\documentclass[aps,prc,twocolumn,tightenlines,superscriptaddress]{revtex4}
\usepackage{epsfig,rotating}
\usepackage{multirow}

\newcommand{\nuc}[2]{\hbox{$^{#1}$#2}}

\begin{document}
\title{Inverse-kinematics one-neutron pickup with fast rare-isotope beams}

\author{A.\ Gade}
   \affiliation{National Superconducting Cyclotron Laboratory,
      Michigan State University, East Lansing, Michigan 48824}
   \affiliation{Department of Physics and Astronomy,
      Michigan State University, East Lansing, Michigan 48824}

\author{J.\,A.\ Tostevin}
    \affiliation{Department of Physics, Faculty of Engineering and
      Physical Sciences, University of Surrey, Guildford, Surrey GU2 7XH,
      United Kingdom}
    \affiliation{Department of Physics, Tokyo Institute of Technology,
2-12-1 Ookayama, Meguro-ku, Tokyo 152-8550, Japan}

\author{T.\ Baugher}
    \affiliation{National Superconducting Cyclotron Laboratory,
      Michigan State University, East Lansing, Michigan 48824}
    \affiliation{Department of Physics and Astronomy,
      Michigan State University, East Lansing, Michigan 48824}

\author{D.\ Bazin}
    \affiliation{National Superconducting Cyclotron Laboratory,
      Michigan State University, East Lansing, Michigan 48824}

\author{B.\,A.\ Brown}
    \affiliation{National Superconducting Cyclotron Laboratory,
      Michigan State University, East Lansing, Michigan 48824}
    \affiliation{Department of Physics and Astronomy,
      Michigan State University, East Lansing, Michigan 48824}

\author{C. M.\ Campbell\footnote{Present Address: Nuclear Science
Division, Lawrence Berkeley National Laboratory, Berkeley,
California 94720, USA.}}
    \affiliation{National Superconducting Cyclotron Laboratory,
      Michigan State University, East Lansing, Michigan 48824}

\author{T.\ Glasmacher}
    \affiliation{National Superconducting Cyclotron Laboratory,
      Michigan State University, East Lansing, Michigan 48824}
    \affiliation{Department of Physics and Astronomy,
      Michigan State University, East Lansing, Michigan 48824}

\author{G. F. Grinyer\footnote{Present Address: GANIL, BP 55027, F-14076
Caen Cedex 5, France.}} \affiliation{National Superconducting
Cyclotron Laboratory,
      Michigan State University, East Lansing, Michigan 48824}

\author{S.\ McDaniel}
    \affiliation{National Superconducting Cyclotron Laboratory,
      Michigan State University, East Lansing, Michigan 48824}
    \affiliation{Department of Physics and Astronomy,
      Michigan State University, East Lansing, Michigan 48824}

\author{K.\ Meierbachtol}
\affiliation{National Superconducting Cyclotron Laboratory,
      Michigan State University, East Lansing, Michigan 48824}
\affiliation{Department of Chemistry, Michigan State University, East Lansing,
  Michigan 48824, USA}

\author{A.\ Ratkiewicz}
    \affiliation{National Superconducting Cyclotron Laboratory,
      Michigan State University, East Lansing, Michigan 48824}
    \affiliation{Department of Physics and Astronomy,
      Michigan State University, East Lansing, Michigan 48824}

\author{S. R.\ Stroberg}
    \affiliation{National Superconducting Cyclotron Laboratory,
      Michigan State University, East Lansing, Michigan 48824}
    \affiliation{Department of Physics and Astronomy,
      Michigan State University, East Lansing, Michigan 48824}

\author{K. A.\ Walsh}
    \affiliation{Department of Physics and Astronomy,
      Michigan State University, East Lansing, Michigan 48824}

\author{D.\ Weisshaar}
    \affiliation{National Superconducting Cyclotron Laboratory,
      Michigan State University, East Lansing, Michigan 48824}

\author{R.\ Winkler}
    \affiliation{National Superconducting Cyclotron Laboratory,
      Michigan State University, East Lansing, Michigan 48824}

\date{\today}

\begin{abstract}
New measurements and reaction model calculations are reported for
single neutron pickup reactions onto a fast \nuc{22}{Mg} secondary
beam at 84 MeV per nucleon. Measurements were made on both carbon
and beryllium targets, having very different structures, allowing a
first investigation of the likely nature of the pickup reaction mechanism. The
measurements involve thick reaction targets and $\gamma$-ray
spectroscopy of the projectile-like reaction residue for final-state
resolution, that permit experiments with low incident beam rates
compared to traditional low-energy transfer reactions. From measured
longitudinal momentum distributions we show that the $\nuc{12}{C}
(\nuc{22}{Mg},\nuc{23}{Mg}+\gamma)X$ reaction largely proceeds as a
direct two-body reaction, the neutron transfer producing bound \nuc{11}{C} target residues.
The corresponding reaction on the \nuc{9}{Be} target seems to largely
leave the \nuc{8}{Be} residual nucleus unbound at excitation energies
high in the continuum. We discuss the possible use of such fast-beam
one-neutron pickup reactions to track single-particle strength in
exotic nuclei, and also their expected sensitivity to neutron
high-$\ell$ (intruder) states which are often direct indicators of
shell evolution and the disappearance of magic numbers in the exotic
regime.
\end{abstract}

\maketitle

\section{Introduction\label{intro}}

The spectroscopy and ordering of nucleon single-particle levels
along extended isotopic chains is of importance for understanding
emerging and dissolving shell structures. One- and two-nucleon
removal reactions from fast rare-isotope beams are making a
significant contribution to such studies in some of the rarest
isotopes~\cite{Han03,Gad07,Adr08}. This information, in turn, allows
an assessment of shell model effective interactions and of their
predictions near both the weakly- and strongly-bound Fermi-surfaces
in highly neutron-proton asymmetric nuclei. By their nature these
nucleon removal reactions preferentially populate states in
the heavy residual nuclei with a strong hole-like parentage upon the
projectile ground state. For the spectroscopy of particle-like
states, light-ion single-nucleon transfer reactions, such as the
$(d,p)$ reaction, are very often the reaction of choice. These
reactions are best and most often carried out at relatively low
incident energies such that linear and angular momentum matching and
hence the reaction yields are optimal. The beam intensity, (thin)
target, and detection system demands for such measurements, of
final-state-resolved transfer cross section angular distributions in
inverse kinematics, are however high and such studies remain
impractical, currently, for many of the most exotic nuclei.

In this paper we consider test case measurements and associated
direct reaction model calculations for reaction events in which a
single neutron is picked
up by a fast secondary beam of mass $A$. The measurements employ
thick targets and $\gamma$-ray spectroscopy and thus take full
advantage of fast beams produced by fragmentation; the resulting
high luminosity allows for experiments with low incident beam
rates~\cite{Gad08}. Related previous studies \cite{pickuppp,al23}
considered reactions involving the pickup of a strongly bound proton
from a \nuc{9}Be target. Here we discuss first measurements for a
\nuc{22}Mg beam of about 84~MeV per nucleon incident energy. An
important aspect of the present analysis is that measurements are made
using both carbon and beryllium reaction targets, with very
different neutron structures. Two of our primary aims here are to
investigate (a) the nature of the fast pickup reaction mechanism,
and (b) the magnitudes of the measured and calculated pickup reaction
yields. In doing so we hope to gain a first insight into the potential
to use such reactions, in combination with $\gamma$-ray spectroscopy
of the populated mass $A+1$ final states, to determine quantitative particle
state spectroscopy information on exotic nuclei produced at energies
of order 100~MeV per nucleon. Specifically, we aim to clarify if the
reaction proceeds predominantly by a direct single-particle pickup
mechanism and, if this is the case, to quantify the measured and
calculated cross sections for transfers involving different
orbital angular momenta $\ell$. Such information will illustrate,
for instance, the capability for future measurements to identify
high-$\ell$ neutron intruder components in the low energy spectra
of the pickup residues. Since such high-$\ell$ intruder configurations
are typically angular momentum mismatched, they are more weakly coupled
and populated in lower energy $(d,p)$ transfer reactions.

We make clear in advance that the fast nucleon pickup events
considered here are not well-matched in either transferred linear or
angular momentum, in the sense used in semi-classical model discussions
of transfer reactions between heavy ions~\cite{Brink72,Phil77}. We
also make clear that the $^{23}$Mg residues in the present analysis,
having a high first nucleon threshold, do not permit a detailed or
quantitative spectroscopy in this case. Our aims here are more modest
and were stated above. In addition, to help direct future experiments,
we assess the effectiveness of our two light target choices: (a)
$^{12}$C, that makes available four well-bound $p_{3/2}$ neutrons
with a ground-state separation energy of $S_n =18.72$~MeV, and (b)
$^{9}$Be, that offers one weakly-bound valence neutron of
$S_n=1.665$~MeV, while $S_{2n}> 20$~MeV, in providing a source of
neutrons with sufficiently high momentum components to contribute
strength to the pickup reaction yields. We will show that the
measured and calculated pickup reaction yields and the $^{23}$Mg
residue momentum distributions measured on the two targets help to
provide such a clarification and guide to future studies.

\section{Experimental Details}

The projectile beam of \nuc{22}{Mg} was obtained by fragmentation of
a 170-MeV per nucleon \nuc{24}{Mg} primary beam provided by the
Coupled Cyclotron Facility at the National Superconducting Cyclotron
Laboratory (NSCL) on the campus of Michigan State University. The
\nuc{9}{Be} fragmentation target of 1904~mg/cm$^2$ thickness was
located at the midacceptance target position of the A1900 fragment
separator~\cite{a1900}. An achromatic aluminum wedge degrader of
600-mg/cm$^2$ thickness and slit systems were used to purify the
beam. The momentum acceptance of the separator was restricted to
$\Delta p /p=0.14\%$. The resulting cocktail beam of $N=10$ isotones
contained \nuc{22}{Mg} (74\%), \nuc{21}{Na} (24.5\%) and
\nuc{20}{Ne} (1.5\%).

Targets of 188-mg/cm$^2$ thick \nuc{9}{Be} and 149.4-mg/cm$^2$ thick
\nuc{nat}{C} (vitreous Carbon with a density of $\rho= 1.54$
~g/cm$^3$, 98.9\% enriched in \nuc{12}{C}) were placed at the
reaction target position of the S800 spectrograph~\cite{s800} to
induce the one-neutron pickup onto the \nuc{22}{Mg} projectiles. The
target position was surrounded by the high-resolution $\gamma$-ray
detection system SeGA, an array of 32-fold segmented high-purity
germanium detectors~\cite{sega}. The high degree of segmentation
allows for event-by-event Doppler reconstruction of the $\gamma$
rays emitted by the projectile-like reaction residues in flight. The
angle of the $\gamma$-ray emission entering the Doppler
reconstruction is determined from the location of the segment that
registered the largest energy deposition. The photopeak efficiency
of the detector array was calibrated with standard sources and
corrected {\it in-beam} for the Lorentz boost of the $\gamma$-ray
distribution emitted by nuclei moving at $v/c > 0.35$.

\begin{figure}[h]
\includegraphics[width=75mm]{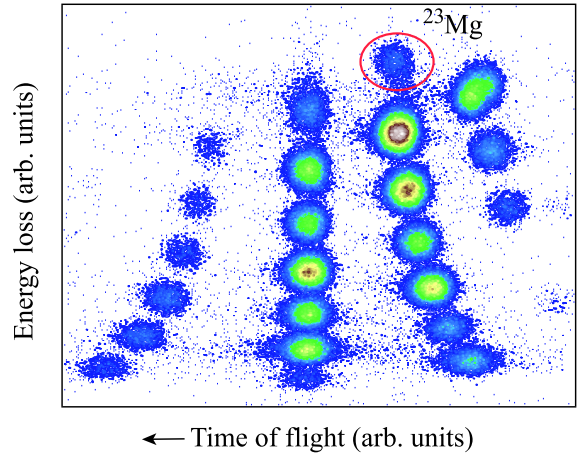}
\caption{\label{fig:pid} (Color online) Particle identification
spectrum (energy loss vs. time of flight) for \nuc{23}{Mg} produced
in the $\nuc{12}{C}(\nuc{22}{Mg},\nuc{23}{Mg}+\gamma)X$ one-neutron
pickup reaction. The spectrum shows all projectile-like reaction
residues from the interaction of \nuc{22}{Mg} with the C target that
entered the S800 focal plane in the same magnetic rigidity setting
as \nuc{23}{Mg}.}
\end{figure}

Event-by-event particle identification was performed with the
focal-plane detection system~\cite{FPdets} of the large acceptance
S800 spectrograph. The energy loss measured with the S800 ionization
chamber and time-of-flight information taken between plastic
scintillators, corrected for the angle and momentum of each ion,
were used to unambiguously identify the projectile-like reaction
residues exiting the target. The incoming projectiles were
identified from their flight-time difference measured with plastic
timing detectors. The particle-identification spectrum for
\nuc{23}{Mg} produced in \nuc{22}{Mg} + \nuc{12}{C} is shown in
Fig.~\ref{fig:pid}.

\begin{figure}[h]
\includegraphics[width=85mm]{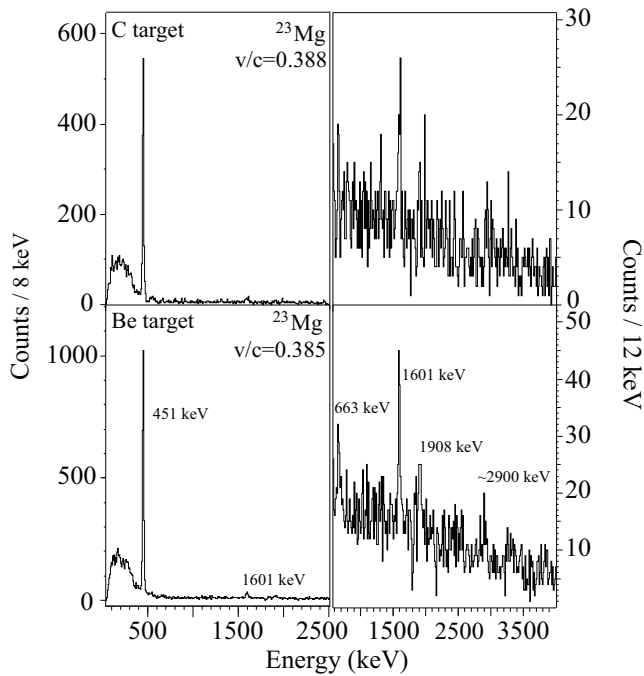}
\caption{\label{fig:gamma} Event-by-event Doppler reconstructed
$\gamma$-ray spectra measured in coincidence with \nuc{23}{Mg}
populated in $\nuc{12}{C}(\nuc{22}{Mg},\nuc{23}{Mg}+\gamma)X$ (top)
and $\nuc{9}{Be}(\nuc{22}{Mg},\nuc{23}{Mg}+\gamma)X$ (bottom),
respectively. Left: The prominent $\gamma$-ray transition at 451~keV
depopulates the $5/2_1^+$ excited state to the $3/2^+$ ground state;
right: expansion of the higher energy part of the $\gamma$-ray
energy spectrum. The other $\gamma$ rays correspond to transitions
from the known $7/2^+$, $1/2^+$, and $9/2^+$ or $5/2^+$ excited
states at 2052, 2359, and 2715~keV~\cite{nndc}, respectively. There
is some indication of a transition at about 2900~keV, which may
depopulate the alleged $(3/2,5/2)^+$ state at 2908~keV excitation
energy~\cite{nndc}.}
\end{figure}

For each target, the inclusive cross section for the one-neutron
pickup to all bound final states of \nuc{23}{Mg} was determined from
the yield of detected pickup residues divided by the number of
incoming projectiles relative to the number density of the
\nuc{9}{Be} and \nuc{12}{C} reaction targets, respectively. The
$\gamma$-ray spectra observed in coincidence with \nuc{23}{Mg},
event-by-event Doppler reconstructed, are displayed in
Fig.~\ref{fig:gamma}. Partial cross sections to individual final
states were obtained from the efficiency-corrected full-energy peak
areas relative to the number of pickup products and corrected for
feeding. Unfortunately, the statistics were not sufficient to tag
the final state of the \nuc{11}{C} target nuclei in the
laboratory-frame $\gamma$-ray spectra $(v/c \approx 0)$.

\begin{figure}[h]
\includegraphics[width=80mm]{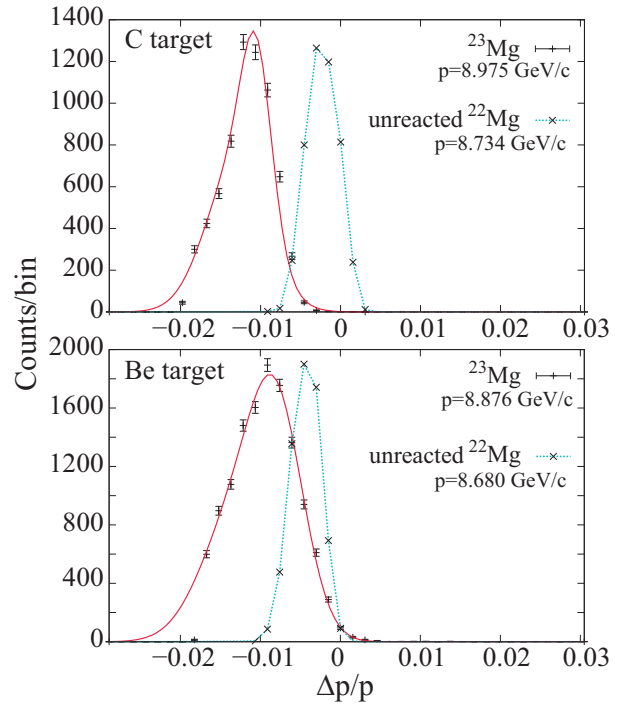}
\caption{\label{fig:momentum} (Color online) Momentum distributions
of the one-neutron pickup product \nuc{23}{Mg} and momentum profile
of the \nuc{22}{Mg} projectile beam passing through the Be and C
target, respectively. The lines are to guide the eye and have been
used for the \nuc{23}{Mg} distribution to estimate acceptance losses
on the low-momentum side caused by a beam blocker in the S800 focal
plane that was necessary to prevent the direct beam from entering
the detection system.}
\end{figure}

\begin{figure}[h]
\includegraphics[width=80mm]{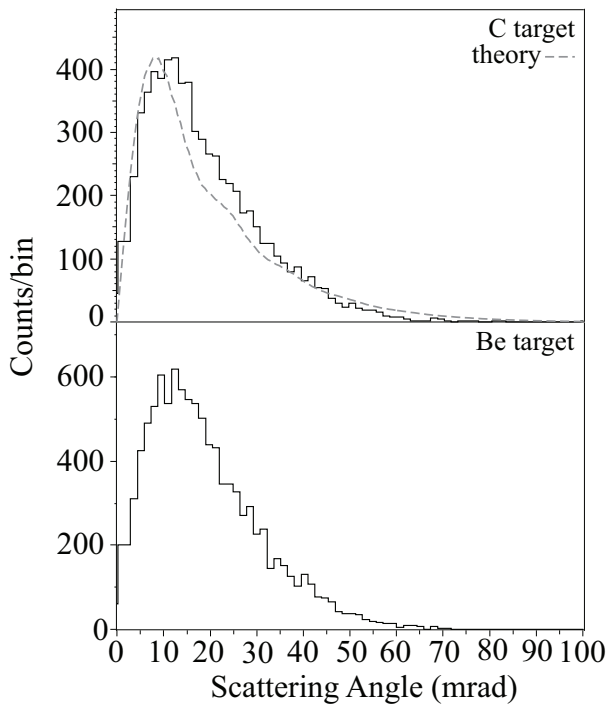}
\caption{\label{fig:scatter} Laboratory frame scattering-angle
distribution, $d\sigma/d\theta_L$, of the heavy residues measured in
the experiment for the carbon (upper panel) and the beryllium (lower
panel) targets. For the carbon target a calculated laboratory frame
angular distribution is also shown (dashed curve) for pickup to a
single final state, the 0.451~MeV, $5/2^+$ state (with \nuc{11}{C}
left in its $3/2^-$ ground state). The latter curve has been scaled to
the measurements to aid comparison of the shapes but does not take
into account small folding effects due to the finite emittance
of the incident beam and from angular straggling in the target. The
spectra show that the distributions fit well into the elliptical
angular acceptance of the S800 spectrograph ($\pm$ 5$^{\circ}$
$\times$ $\pm$ 3.5$^{\circ}$) in agreement with the calculations.}
\end{figure}

Position information from the two cathode readout drift chambers
(CRDCs) of the S800 focal-plane detection system together with
trajectory reconstruction employing the optics code COSY~\cite{cosy}
were used to reconstruct the longitudinal momentum distributions of
the pickup residues on an event-by-event basis.
Fig~\ref{fig:momentum} shows for both the Be and the C targets the
momentum distributions of the pickup product \nuc{23}{Mg} together
with the momentum profile of the \nuc{22}{Mg} beam passing through
the target essentially un-reacted. The width of the \nuc{22}{Mg}
momentum profile as measured in the spectrograph's focal plane is
dominated by the energy (momentum) straggling of the projectiles in
the respective reaction target. The curves through the data points
are to guide the eye and have been used in the case of the
\nuc{23}{Mg} distributions to determine acceptance corrections on
the low-momentum side of 9.9\% and 5\% for the cross sections
measured with the Be and C targets, respectively. The widths of the
\nuc{23}{Mg} momentum distributions are strikingly different for the
measurements with the two different targets while the momentum
profiles of the un-reacted \nuc{22}{Mg} projectiles are essentially
identical in both cases. The differential momentum loss $\delta p$
of \nuc{23}{Mg} that takes into account that the pickup can, with
equal probabilities, occur at any point along the trajectory through
the target was found to be negligible $(\delta p/p < 1.8 \times
10^{-3})$ unlike in the case of one-proton pickup~\cite{pickuppp}.
Figure~ \ref{fig:scatter} shows the measured angular distributions
of the heavy residues as a function of their laboratory frame angle,
$d\sigma/d\theta_L$, for the pickup reactions induced by both
targets. The spectra demonstrate that essentially the entire final
state distributions and yield are within the spectrograph's angular
acceptance.

If the neutron pickup reaction is strictly {\em two-body}, i.e. it involves
only transfers between bound states of the target and the projectile and
also leaves a bound target-like residue, then the intrinsic,
reaction-mechanism-induced parallel momentum distribution of the
heavy projectile-like residues, $d\sigma/dp_\parallel$, will be very
narrow. The necessary formula is given by Eq.\ (1) of Ref.
\cite{pickuppp}. So, unlike knockout reaction measurements, the
residue momentum distribution does not provide spectroscopic (orbital
angular momentum) information, however, it carries a reaction mechanism
diagnostic, as follows.

Following strict two-body transfers, the intrinsic $d\sigma/dp_\parallel$
momentum distribution is essentially $\delta$-function like in comparison
to the incident beam momentum resolution $\Delta p/p \approx 0.14 \%$ and
the energy broadening due to passage through the target. Thus, the extent
to which the measured \nuc{23}{Mg} residue momentum distribution differs
from that of the un-reacted $^{22}$Mg beam provides direct evidence for
events that go beyond a two-body reaction model description. We note that
this direct comparison of the measured \nuc{23}{Mg} distributions from the
carbon and beryllium targets with that of the un-reacted $^{22}$Mg, shown in
Fig. \ref{fig:momentum}, leads to qualitatively different outcomes in the
two cases. We return to this important point in discussions of the calculated
and measured cross sections.

\section{Reaction model calculations}

We consider the neutron-pickup reactions to be described by a post
form, fully-finite-range, coupled channels Born approximation (CCBA)
reaction analysis. The neutron pickup from the light target(s) is
assumed to take place in a single step and we include transfers to
$^{22}$Mg in its ground and first excited $2^+$ state. We carry out
these CCBA calculations using the flexible direct reactions code {\sc
fresco} \cite{fresco}.

\begin{figure}[h]
\bigskip
\includegraphics[width=80mm]{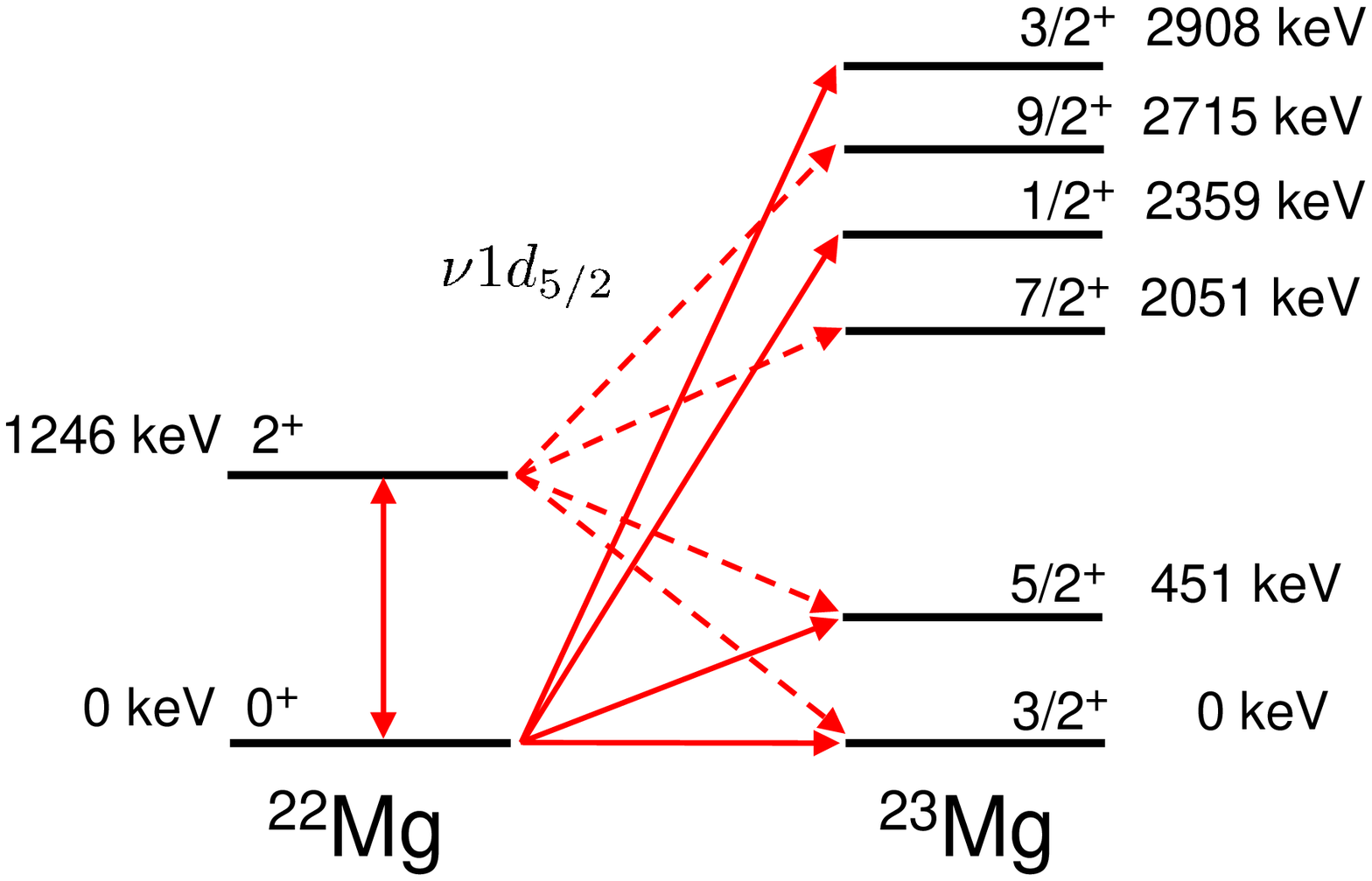}
\caption{\label{fig:reaction} (Color online) Schematic of the CCBA
channel-coupling scheme used for the $^{22}$Mg($^{12} $C,$^{11} $C)$
^{23}$Mg($J^\pi$) and $^{22}$Mg($^{9}$Be,$^{8}$Be*)$^{23}$Mg($ J^\pi
$) neutron pickup reactions. The solid lines (from $^{22}$Mg to $^{
23}$Mg) represent one-step single particle transfers to the final
states indicated. The dashed lines show two-step pathways to $^{23}
$Mg states, by neutron capture onto the $^{22}$Mg$(2^+)$ core. All
of these involve a $1d_{5/2}$ neutron orbital.}
\end{figure}

The (all-order) inelastically coupled $0^+$ and 2$^+$ $^{22}$Mg core
states and the single-step transfer channel paths included in our
calculations are shown in Fig. \ref{fig:reaction}. The $^{23}$Mg
bound states populated by neutron capture onto the $^{22}$Mg$(2^+)$
core, which all involve a $1d_{5/2}$ orbital, are represented by
dashed lines. We note that the proton threshold in $^{23}$Mg is at
$S_p =7.58$ MeV, and thus only the strongest low-lying shell model
states have been included in Fig. \ref{fig:reaction}. The shell
model also predicts fragmented $sd$-shell strength to states up to
the proton threshold that will be quantified, approximately, later.
The entrance and exit channel distorting optical model interactions
used and our treatment of the $^{22}$Mg projectile excitation are
also discussed later.

We first consider the possible implications of the light target nucleus
structures on our reaction treatment. We note that our two-body CCBA
reaction methodology assumes kinematics and dynamics in which the projectile
and target in the initial and final states separate as two bound
systems. The neutron is assumed transferred from bound states in the
target to bound states in the projectile residue, the target residue
remaining bound. For the carbon
target case our expectation is that this two-body picture will be an
excellent approximation (for either neutron or proton pickup). Here
we make use of the $^{12}$C (=$^{11}$C+n) structure information as
was deduced for the corresponding neutron knockout reactions from
$^{12}$C, as discussed in Ref.\ \cite{BHS02}. The ground-state
neutron separation energy is $S_n =18.72$ MeV and neutron pickup
will leave the $^{11}$C residue predominantly in its ground state
(3/2$^-$) or the excited states at 2.000 (1/2$^-$) and 4.804 MeV
(3/2$^-$). These three states lie well below the first $^{11}$C
threshold of 8.69 MeV. Theoretically the shell-model spectroscopic
factors to these three states, calculated with the Warburton and
Brown $p$-shell ({\sc wbp}) interaction \cite{wbp}, are $C^2S=3.16$,
0.58, and 0.19 and are seen to essentially exhaust the four units of
single particle strength expected. The very small remainder, of
0.07, is fragmented over numerous $^{11}$C states above 10 MeV in
excitation. These {\sc wbp} interaction spectroscopic factors agree
well with other $p$-shell shell model calculations and thus this
description of the $^{12}$C target is robust.

We also take from Ref.\ \cite{BHS02} the $^{11}$C+n Woods-Saxon
binding potential geometry used to calculate these $p$-shell neutron
overlaps. These have radius and diffuseness parameters of 1.310 fm
and 0.55 fm, respectively. A $^{11}$C ($^{12}$C) root mean squared
(rms) matter radius of 2.12 (2.32) fm was assumed for the
calculation of the $^{11} $C+$^{23}$Mg ($^{12}$C+$^{22}$Mg)
distorting potentials, as described below.

For the $^{9}$Be target the situation is significantly more complex.
The configurations of the single weakly-bound valence neutron, with
$S_n$=1.665 MeV, relative to the unbound (but near the two-$\alpha$
particle threshold) $^{8}$Be 0$^+$ and 2$^+$ states have been
studied in some detail. The associated neutron overlaps and their
spectroscopic factors from the variational Monte Carlo (VMC) wave
functions of Wiringa {\em et al.} \cite{xx,xxx} and from extended
microscopic cluster model wave functions, e.g. Arai {\em et al.}
\cite{yy}, are in rather close agreement. In both cases the [$0^+
\otimes p_{3/2}]_{3/2}$ and [$2^+ \otimes p_{3/2}]_{3/2}$
configurations are completely dominant. The cluster model (VMC)
spectroscopic factors for these two configurations are 0.553 (0.591)
and 0.514 (0.583),
respectively. In the following we use the numerical $^8$Be $0^+$ and
$2^+$ core state overlaps from Arai {\em et al.} \cite{zz} (as are
shown in Fig. 4(a) of Ref.\ \cite{yy}).

For convenience, these overlaps were fitted to single-particle wave
functions calculated in Woods-Saxon potential wells, with separation
energies of 1.665 and 4.695 MeV. The fitted (reduced radius,
diffuseness) spin-orbit strength parameters were (1.09 fm, 0.59 fm)
6.0 MeV and (1.09 fm, 0.75 fm) 6.0 MeV. As will be discussed further,
the amplitudes for valence neutron pickup will leave the target-like
residues unbound/resonant (though only very weakly so). However, if
the reaction proceeds by the pickup of a more strongly bound neutron
(from the $^8$Be core) the resulting target residues will be left at
high excitation in the continuum; this core state single-particle
strength being in the vicinity of 15 MeV of excitation in
$^8$Be$^*$. We noted already the qualitative difference between the
measured momentum distributions of the $^{23}$Mg residues from reactions
on the beryllium and carbon targets. These, and the cross sections
presented below indicate that such core neutron pickup events are
dominant in the case of the $^9$Be target and hence that our two-body
dynamical model is not appropriate for a quantitative discussion of
fast pickup reaction yields in this case.

Concentrating first on the $^{12}$C case, the pickup reaction is
computed as $^{22}$Mg($^{12}$C,$^{11}$C($I^\pi$))$^{23}$Mg($J^\pi$)
leading to the $I^\pi = 3/2_1^-,\,1/2^-$ and $3/2_2^-$ states of
$^{11}$C at 84 MeV per nucleon incident energy. The (absorptive)
nuclear optical interactions were calculated, as is done in the fast
nucleon knockout reaction studies, by double folding the neutron and
proton densities of $^{22}$Mg ($^{23}$Mg), obtained from spherical
Skyrme (SkX interaction) Hartree-Fock (HF) calculations
\cite{brown98}, and of $^{12}$C ($^{11}$C), assumed a Gaussian with
rms radius given earlier, with an effective nucleon-nucleon (NN)
interaction \cite{Tos04}. The $^{22}$Mg was allowed to inelastically
excite, Fig. \ref{fig:reaction}, enabled by deforming the entrance
channel nuclear distorting
potential with a deformation length of $\delta_2 = 1.95$ fm. This
corresponds to an assumed mass $\beta_2$ value of 0.58, consistent
with the charge $\beta_2$ of $0.58(11)$ from Ref.\ \cite{Raman}. The
binding geometry and spectroscopic factors of the $^{11}$C+n
overlaps were already detailed above.

The required neutron-projectile bound states/overlaps [$^{22}$Mg$
(0^+,\,2^+) \otimes n \ell_j ]_{J}$ and their spectroscopic
amplitudes were taken from full $sd$-shell-model calculations that
use the recently re-derived {\sc usdb}, Universal $sd$-shell
Hamiltonian of Brown and Richter \cite{usdb}. As is indicated in
Fig.\ \ref{fig:reaction}, there are interfering paths for population
of the $^{23}$Mg($3/2^+$) ground state, via the [$ 0^+ \otimes
1d_{3/2}]$ and [$2^+ \otimes 1d_{5/2}]$ transfers. Both of these
paths are weak in the present case resulting in a small predicted
$^{23}$Mg ground state cross section. The first excited $5/2^+$, 451
keV state is also shown to proceed by both direct and indirect,
$2^+$ state, paths. However, this calculated [$2^+ \otimes
1d_{5/2}]$ {\sc usdb} spectroscopic factor is 0.005, so this
(negligible) two-step path is not considered further. The state at
2.715 MeV, shown in the literature as $9/2^+, \, (5/2^+)$, is
assumed in our calculations to be a $9/2^+$ state, and can be
associated with a large amplitude $[2^+ \otimes 1d_{5/2}]_{9/2}$
core-coupled shell model state at 2.762 MeV. There is no shell model
candidate for a $5/2^+$ state near this energy.

The {\sc usdb} shell model spectroscopic factors $C^2S$ are
collected in Tables \ref{tab:theo1} and \ref{tab:theo2}. The
associated single-particle states were calculated in real
Woods-Saxon potential wells, all with diffuseness parameter
$a_0=0.7$ fm and a spin-orbit interaction of strength 6 MeV. The
reduced radius parameters $r_0$ of these potentials were adjusted to
reproduce the rms radius of each single-particle orbital, as given
by the same Hartree Fock calculations as were used to calculate the
projectile densities, as has been discussed in detail elsewhere
\cite{reduction}. These $r_0$ were 1.284, 1.315 and 1.134 fm for the
$1d_{5/2}$, $1d_{3/2}$ and $2s_{1/2}$ orbitals, respectively. These
overlap functions were calculated using their physical separation
energies, computed from the ground state separation energy, $S_n =
13.15$ MeV and the excitation energies. The theoretical reaction
yields for the carbon target are collected in Table \ref{tab:theo1}.
These are shown summed with respect to the three $^{11}$C final
states and which are dominated, because of the large spectroscopic
factor, by the ground state contribution. The cross sections are
computed by integrating the calculated center-of-mass frame
differential cross sections for angles $\theta_{cm}<40$ degrees. At
this upper angle limit the cross sections have fallen to $10^{-9}$
of their values at forward angles.

\begin{table}[t]
\begin{center}
\vspace{0.5cm} \caption{\label{tab:theo1} Experimental cross
sections are compared to the CCBA reaction and shell-model
calculations for the fast single-neutron pickup reactions
$^{22}$Mg($^{12}$C,$^{11}$C($ I^\pi$))$^{23}$Mg($J^\pi$) at 84 MeV per
nucleon. The theoretical cross section $\sigma^{th}$ is inclusive
with respect to the population of the $I^\pi = 3/2^-,\ 1/2^-$ and
$3/2^-$ states of the $^{11}$C target residue. The spectroscopic
amplitudes of the interfering [$0^+ \otimes 1d_{3/2}]$ and [$2^+
\otimes 1d_{5/2}]$ ground state paths have the same sign.}
\begin{ruledtabular}
\begin{tabular}{cccccccc}
$J^{\pi}$ & $E$ &$\sigma_f$ & Configuration&$C^2S$ &\ \ $\sigma^{th}$& \\
&  &       expt.& SM & SM & \\
&(keV)&  (mb)     &              &       &(mb) &\\\hline
3/2$^+$ &0.0&$\leq 0.77^{+0.09}_{-0.13}$ & [$0^+ \otimes 1d_{3/2}]$ & 0.054 & 0.083 &  \\
&   &                       & [$2^+ \otimes 1d_{5/2}]$ & 0.939 &  &  \\
5/2$^+$ &451&1.27(14)               & [$0^+ \otimes 1d_{5/2}]$ & 0.410 & 1.448 &  \\
7/2$^+$ &2051&0.18(5)               & [$2^+ \otimes 1d_{5/2}]$ & 0.574 & 0.054 &  \\
1/2$^+$ &2359&0.08(5)               & [$0^+ \otimes 2s_{1/2}]$ & 0.242 & 0.010 &  \\
9/2$^+$ &2715&0.10(5)               & [$2^+ \otimes 1d_{5/2}]$ & 0.366 & 0.096 &  \\
3/2$^+$ &2908& ---                & [$0^+ \otimes 1d_{3/2}]$   & 0.238 & 0.326 &  \\
\hline \multicolumn{7}{c}{inclusive cross section: 2.40(19)~mb}\\
\end{tabular}
\end{ruledtabular}
\end{center}
\end{table}

Calculations in the $^{9}$Be target case were identical as far as
the projectile-like system is concerned. Regarding the target, the
neutron pickup was computed as due only to the weakly bound valence
neutron, i.e. $^{22}$Mg($^{9}$Be,$^{8}$Be($ 0^+, \, 2^+ $))$
^{23}$Mg($J^\pi$). The summed spectroscopic strength to these two
$^{8}$Be final states is of order unity. The nuclear distorting
interactions were calculated assuming Gaussian $^{9}$Be and $^{8}$Be
matter densities both with rms radius of 2.36 fm. The $^{22}$Mg
deformation was treated as for the carbon target case. The binding
geometry and spectroscopic factors of the $^{8}$Be+n overlaps were
detailed above, taken from the model of Arai {\em et al.} \cite{yy}.
The theoretical reaction yields are collected in Table
\ref{tab:theo2}, where they are summed with respect to the two
$^{8}$Be final states.
\begin{table}[h]
\begin{center}
\vspace{0.5cm} \caption{\label{tab:theo2} Experimental cross
sections are compared to the CCBA reaction and shell-model
calculations for the fast single-neutron pickup reactions
$^{22}$Mg($^{9}$Be,$^{8}$Be($ I^\pi$))$^{23}$Mg($J^\pi$) at 84 MeV per
nucleon. The theoretical cross section $\sigma^{th}$ is inclusive
with respect to the population of the $I^\pi = 0^+$ and $2^+$ states
of the $^{8}$Be target residue. The spectroscopic amplitudes of the
interfering $[0^+ \otimes 1d_{3/2}]$ and $[2^+ \otimes 1d_{5/2}]$
ground state paths have the same sign.}
\begin{ruledtabular}
\begin{tabular}{cccccccc}
$J^{\pi}$ & $E$ &$\sigma_f$ & Configuration&$C^2S$ &\ \ $\sigma^{th}$& \\
&  &       expt.& SM & SM & \\
&(keV)&  (mb)     &              &       &(mb) &\\\hline
3/2$^+$ &0.0&$\leq 0.86^{+0.08}_{-0.11}$ & [$0^+ \otimes 1d_{3/2}]$ & 0.054 & 0.010 &  \\
&   &                       & [$2^+ \otimes 1d_{5/2}]$ & 0.939 &       &  \\
5/2$^+$ &451&1.32(12)               & [$0^+ \otimes 1d_{5/2}]$ & 0.410 & 0.206 &  \\
7/2$^+$ &2051&0.15(4)               & [$2^+ \otimes 1d_{5/2}]$ & 0.574 & 0.006 &  \\
1/2$^+$ &2359&0.13(4)               & [$0^+ \otimes 2s_{1/2}]$ & 0.242 & 0.003 &  \\
9/2$^+$ &2715&0.13(4)               & [$2^+ \otimes 1d_{5/2}]$ & 0.366 & 0.010 &  \\
3/2$^+$ &2908& ---                & [$0^+ \otimes 1d_{3/2}]$   & 0.238 & 0.037 &  \\
\hline \multicolumn{7}{c}{inclusive cross section: 2.58(16)~mb}\\
\end{tabular}
\end{ruledtabular}
\end{center}
\end{table}

\section{Results and discussion}

The $\nuc{12}{C}(\nuc{22}{Mg},\nuc{23}{Mg}+\gamma)X$ one-neutron
pickup reaction was performed at 84.7~MeV per nucleon mid-target
energy. The longitudinal momentum distribution of the \nuc{23}{Mg}
residues, reconstructed in the focal plane of the S800 spectrograph
[see Fig.~\ref{fig:momentum} (top)], was cut by the spectrograph's
beam blocker; necessary to stop the direct \nuc{22}{Mg} beam passing
through the target. A 5\% correction for the missing counts was
applied to the cross section. The inclusive cross section, including
the acceptance correction, amounts to $\sigma_{inc}=2.40(19)$~mb. A
6\% systematic uncertainty, attributed to fluctuations in the
incoming beam composition, was added in quadrature to the
statistical uncertainty. From the $\gamma$-ray spectra taken in
coincidence with the \nuc{23}{Mg} reaction products, the partial
cross sections for the population of the $5/2_1^+$, $7/2_1^+$,
$1/2_1^+$ and the proposed $9/2^+,5/2^+$ level at 451, 2052, 2359
and 2715~keV, respectively, were obtained from the intensity of the
observed $\gamma$-ray transitions and the known feeding patterns.
Known transitions that were not observed in the present experiment
due to limited statistics or $\gamma$-ray detection efficiency, for
example the 2359 keV ground-state transition of the first $1/2^+$
state, are included in our partial cross sections by using the
reported branching ratios in the literature~\cite{nndc}. The partial
cross section for the $3/2^+$ ground state was derived from the
inclusive cross section by subtraction of all observed feeders. The
possible population of the $(3/2,5/2)^+$ state at 2.9~MeV is
included in the uncertainty. We stress that, due to possible
unobserved feeding by higher energy $\gamma$-ray transitions, the
ground state cross section, in particular, should be considered an
upper limit. The measured cross sections are summarized in
Table~\ref{tab:theo1} where they are compared to calculations.

The results for the calculation of the carbon-induced reaction are
expected to be reliable, quantitatively. The momentum distribution
comparison made in Fig. \ref{fig:momentum} provides first evidence
that the reaction proceeds as a direct neutron transfer reaction
producing {\em bound} $^{11}$C residues. The calculated cross sections
track the values and trends of the measurements reasonably
accurately, the exception being the larger experimental yield
recorded against the ground state transition (i.e. all reaction
events without an identifiable $\gamma$ ray). As was stated earlier,
the present $^{23}$Mg final state case is very challenging, with the
proton decay threshold only at 7.58 MeV. As will be seen from Table
\ref{tab:theo1}, the {\sc usdb} shell model calculation predicts a state with
appreciable [$0^+ \otimes 1d_{3/2}]$ strength near a known $(5/2, \,
3/2)^+$ state at 2.908 MeV with a calculated cross section of 0.33
mb. This state was not clearly identified in the current experiment;
however, assuming that the peak structure in the $\gamma$-ray
spectra at 2900~keV is the ground-state transition of this state,
the partial cross section for its population is estimated to be around 0.2~mb.
Similarly, the shell model predicts a summed [$0^+ \otimes
1d_{5/2}]$ strength of 0.64 to final states up to 8 MeV in excitation in
$^{23}$Mg. Only 0.41 of this strength has been accounted for in the
451 keV $5/2^+$ state, that is responsible for a calculated cross
section of 1.45 mb. It is likely therefore that there is population
of fragmented $d$-wave states with a summed direct reaction cross
sections yield near to that recorded against the ground state,
$0.77^{+0.09}_{-0.13}$ mb. This value should thus be considered as
an upper limit and an estimate of these direct populations of
unobserved fragmented states and that decay by energetic $\gamma$
rays. In the present case, in addition to the $sd$-shell model
states discussed, there may also be unobserved contributions from
bound $f_{7/2}$ fragments at higher excitation energy. The present
experiment and data set shows no evidence for and does not allow any
meaningful discussion of such configurations.

The $\nuc{9}{Be}(\nuc{22}{Mg},\nuc{23}{Mg}+\gamma)X$ one-neutron
pickup reaction was performed at 84.2~MeV per nucleon mid-target
energy. The inclusive cross section was measured for several data
runs and found to be constant within the statistical uncertainty.
Since for this target the longitudinal momentum distribution of the
\nuc{23}{Mg} residues is much wider than for the carbon target-induced reaction
[see Fig.~\ref{fig:momentum} (bottom)], the low-momentum tail of the
distribution is cut more severely by the beam blocker. A 9.9\% correction
for the missing counts was applied to the cross section. The
inclusive cross section, including the acceptance correction and
with a 6\% systematic uncertainty added, amounts to
$\sigma_{inc}= 2.58(16)$~mb. The $\gamma$-ray spectra and population
patterns for the carbon and beryllium-induced reactions are very similar. In
fact, the partial cross sections are identical within uncertainties.
Again, the partial cross section for the $3/2^+$ ground state was
obtained from subtracting the populations from all observed feeders,
with the possible population of the $(3/2,5/2)^+$ state at 2.9~MeV
included in the uncertainty. In addition, due to possible unobserved
feeding by higher energy $\gamma$-ray transitions, the ground state
cross section is expected to be an upper limit. The measured cross
sections are summarized in Table~\ref{tab:theo2} and discussed below.

The results for the $^{9}$Be target allow us to draw some immediate
conclusions. The cross sections in Table \ref{tab:theo2}, calculated
assuming pickup only of the valence neutron, are considerably smaller
than those measured. This result, combined with the similarity of
the measured values to those of the carbon target, and the
(non-two-body) width of the measured $^{23}$Mg momentum distribution
for this target, see Fig. \ref{fig:momentum}, all point to the
dominance of reaction events involving the pickup of strongly-bound
neutrons from $^{9}$Be. We conclude that the momentum composition of
the wave function of the weakly-bound neutron does not match effectively
to the needs of the fast pickup mechanism. As
was discussed earlier, the strength associated with the core
neutrons is located at $^{8}$Be excitation energies high in the
continuum, near to and in excess of 15 MeV, and so cannot be modelled
quantitatively within the CCBA framework. It is interesting to observe that
empirically these four neutrons in the (two $\alpha$-particle)
$^{8}$Be core contribute approximately the same cross sections,
state-by-state, as for the carbon target (see Table
\ref{tab:theo1}).

It can be deduced from Table \ref{tab:theo1} that the fast
pickup reaction is also selective in populating the states with
higher orbital angular momentum transfer. In the present experiment
our ability to probe this aspect of the reaction is limited since
only $s$- and $d-$wave neutron final states are resolved. The [$0^+
\otimes 1d]$ configurations are seen to dominate. However, we can
begin to clarify this expected orbital angular momentum sensitivity
using our theoretical CCBA calculations for the carbon target. To make
transparent this $\ell$-sensitivity, without the need to remove
spectroscopic factors and $Q$-value considerations, we calculated the
pickup cross sections for assumed
$2s_{1/2}$, $1p_{3/2}$, $1d_{5/2}$, and $1f_{7/2}$ transfers, all
with unit spectroscopic factor and with all states located at the position of
the physical 0.451 MeV state; and so with separation energy 12.70
MeV. The summed (over \nuc{11}{C} final states) single-particle
cross sections obtained are $\sigma_{sp}= 0.04$, 0.58, 3.51, and
11.12 mb, respectively. These values represent real enhancements
with $\ell$, due to improved linear and angular momentum matching,
above those due to the advantageous $2J+1$ final states multiplicative
factor. This is clear from the re-scaled values
$\sigma_{sp}/(2J+1) = 0.02$, 0.116, 0.585, and 1.39~mb,
respectively. This suggests that there is the potential for this
reaction mechanism to help locate emerging particle strength
associated with high-$\ell$ intruder orbitals that enter the
low-energy levels spectrum of rare nuclei.

\section{Concluding comments}

In summary, new measurements and reaction model calculations are
presented for the fast one-neutron pickup reactions
$\nuc{12}{C}(\nuc{22}{Mg},\nuc{23}{Mg}+\gamma)X$ and
$\nuc{9}{Be}(\nuc{22}{Mg},\nuc{23}{Mg}+\gamma)X$ at mid-target
energies of about 84 MeV per nucleon. Measurements were made using
both carbon and beryllium targets, having distinctly different
neutron single-particle configurations. Significant differences in
the widths of the \nuc{23}{Mg} longitudinal momentum distributions
for the two different targets were observed, pointing to the
differences in the corresponding reaction mechanisms. These data
thus provide evidence that the $\nuc{12}{C}(\nuc{22}{Mg},
\nuc{23}{Mg}+\gamma)X$ reactions proceed largely as a direct neutron
transfer - producing \nuc{11}{C} target residues in bound states -
while the corresponding pickup reactions induced by \nuc{9}{Be}
appear to leave the \nuc{8}{Be} target-like residue highly excited
and in the continuum.

Partial cross sections to \nuc{23}{Mg} final states are calculated
(inclusive with respect to the target residue final states) based on
the coupled channels Born approximation and assuming shell-model
configurations for the \nuc{23}{Mg} final states. These cross
sections are in reasonable quantitative agreement with the measured
excited state partial cross section values. Clearly these pickup
partial cross sections in themselves provide insufficient information to determine
empirically both the dominant single-particle transferred angular
momentum and their spectroscopic strengths - and must be used in
conjunction with structure theory. Based on our observed (theoretical)
sensitivity of the fast pickup reactions to high-momentum components in the
nuclear wave function and the $\ell$ of the transferred nucleon, we
propose their possible application in
helping to map the descent of high-$\ell$ neutron single-particle
(intruder) states in regions of shell evolution where traditional
neutron magic numbers break down. Specifically, we envisage that the
fast pickup mechanism could be used, together with theoretical
predictions of level
ordering and spectroscopic factors, to study the systematics of and
test level migration predictions along isotopic chains - assessing
shell-model and effective interaction predictions. This could
provide complementary information to transfer reaction studies,
where these overlap, and with the prospect of extending measurements
into regions currently inaccessible (by virtue of beam intensity) to
transfer. Examples are the most neutron-rich Ne, Na and Mg isotopes
approaching $N=20$ where the neutron intruder $f_{7/2}$ orbital
comes down in energy and dominates ground-state and low-lying
configurations in the so-called ``Island of Inversion''
\cite{War90}.

\begin{acknowledgments}
The authors would like to thank Dr Koji Arai for providing tables of
the overlaps, from Figure 4(a) of Ref. \protect\cite{yy}, expressed
with conventional shell-model spin couplings. This work was
supported by the National Science Foundation under Grants No.
PHY-0606007 and PHY-0758099 and by the UK Science and Technology
Facilities Council (STFC) through Research Grant No. ST/F012012. JAT
gratefully acknowledges the financial and facilities support
of the Department of Physics, Tokyo Institute of Technology.

\end{acknowledgments}

\end{document}